\documentclass[%
 aip,
 amsmath,amssymb,
preprint,%
]{revtex4-1}

\usepackage{graphicx}
\usepackage{dcolumn}
\usepackage{bm}

\usepackage[utf8]{inputenc}
\usepackage[T1]{fontenc}
\usepackage{mathptmx}

\usepackage[T1]{fontenc}
\usepackage[latin9]{luainputenc}
\usepackage{float}
\usepackage{textcomp}
\usepackage{amsmath}
\usepackage{amsthm}
\usepackage{amssymb}
\usepackage{graphicx}

\makeatletter
\theoremstyle{plain}
\newtheorem{thm}{\protect\theoremname}
\theoremstyle{definition}
 \newtheorem{example}{\protect\examplename}
 \newtheorem{prop}[thm]{\protect\propositionname}

\makeatother

\providecommand{\examplename}{Example}
\providecommand{\theoremname}{Theorem}
\providecommand{\propositionname}{Proposition}

\begin{document}

\preprint{AIP/123-QED}

\title[Explicit Unsteady Navier-Stokes Solutions and their Analysis via Local Vortex Criteria]{Explicit Unsteady Navier-Stokes Solutions and their Analysis via Local Vortex Criteria}

\author{Tiemo Pedergnana}
\affiliation{%
Institute of Energy Technology,  ETH Z{\"u}rich, Sonneggstrasse 3,
8092 Z{\"u}rich, Switzerland
}%
\author{David Oettinger}%
\affiliation{ 
ELCA Informatik AG, Steinstrasse 21, 8003 Zürich, Switzerland
}%
\author{Gabriel P. Langlois}%
\affiliation{ 
Division of Applied Mathematics, Brown University, Providence, RI 02912, United States
}%

\author{George Haller}
  \email{georgehaller@ethz.ch}
\affiliation{Institute for Mechanical Systems, ETH Z{\"u}rich Leonhardstrasse 21,
8092 Zürich, Switzerland
}%

\date{\today}

\begin{abstract}
We construct a class of spatially polynomial velocity fields that
are exact solutions of the planar unsteady Navier--Stokes equation.
These solutions can be used as simple benchmarks for testing numerical
methods or verifying the feasibility of flow-feature identification
principles. We use examples from the constructed solution family to
illustrate deficiencies of streamlines-based feature detection and
of the Okubo--Weiss criterion, which is the common two-dimensional
version of the broadly used $Q$-, $\Delta$-, $\lambda_{2}$- and
$\lambda_{ci}$-criteria for vortex-detection. Our planar polynomial
solutions also extend directly to explicit, three-dimensional unsteady
Navier--Stokes solutions with a symmetry.
\end{abstract}

\maketitle

\section{Introduction}

In this paper, we address the following question: For what time-dependent
vectors $\boldsymbol{a}_{kj}(t)\in\mathbb{R}^{2}$ does the two-dimensional
(2D) velocity field 
\begin{equation}
\mathbf{u}(\mathbf{x},t)=\sum_{j=0}^{n}\sum_{k=0}^{m}\mathbf{a}_{kj}(t)x^{k}y^{j}\label{eq: Full velocity field, general form}
\end{equation}
solve the incompressible Navier--Stokes equation with the spatial
variable $\mathbf{x}=(x,y)\in\mathbb{R}^{2}$ and the time variable
$t\in\mathbb{R}$? Answering this question enables one to produce
a large class of exact Navier--Stokes solutions for numerical benchmarking
and for verifying theoretical results on simple, dynamically consistent,
unsteady flow models. For linear velocity fields ($\mathbf{a}_{kj}(t)\equiv0$
for $k,j>1$), general existence conditions are detailed by Majda
\cite{majda1986vorticity} and Majda and Bertozzi \cite{majda2002vorticity}.
A more specific form of these spatially linear solutions is given
by Craik and Criminale \cite{craik1986evolution}, who obtain that
any differentiable function $\boldsymbol{a}_{00}(t)$ and any differentiable,
zero-trace matrix $\mathbf{A}(t)$ generates a linear Navier--Stokes
solution $\mathbf{u}(\mathbf{x},t)$ in the form 
\begin{equation}
\mathbf{u}(\mathbf{x},t)=\mathbf{a}_{00}(t)+\mathbf{A}(t)\mathbf{x},\label{eq: Linear velocity field}
\end{equation}
provided that $\dot{\mathbf{A}}(t)+\mathbf{A}^{2}(t)$ is symmetric.
Note that all such solutions are universal (i.e., independent of the
Reynolds number) because the viscous forces vanish identically on
them. Reviews of exact Navier--Stokes solutions tend to omit a discussion
of the Craik--Criminale solutions, although they list several specific
spatially linear steady solutions for concrete physical settings (see
Berker \cite{berker1963integration}, Wang \cite{wang1989exact,wang1990exact,wang1991exact},
and Drazin and Riley \cite{drazin2006navier}). For solutions with
at least quadratic spatial dependence, no general results of the specific
form (1) have apparently been derived.

In related work, Perry and Chong \cite{perry1986series} outline
a recursive procedure for determining local Taylor expansions of solutions
of the Navier--Stokes equation up to any order when specific boundary
conditions are available. Bewley and Protas \cite{bewley2004skin}
show that near a straight boundary, the resulting Taylor coefficients
can all be expressed as functions and derivatives of the skin friction
and the wall pressure. The objective in these studies is, however,
a recursive construction of Taylor coefficients for given boundary
conditions, rather than a derivation of the general form of exact
polynomial Navier--Stokes velocity fields of finite order. We also
mention the work of Bajer and Moffatt \cite{bajer1990class}, who
construct exact, steady three-dimensional (3D) Navier--Stokes flows
with quadratic spatial dependence for which the flux of the velocity
field through the unit sphere vanishes pointwise.

The linear part of the velocity field (\ref{eq: Full velocity field, general form})
can already have arbitrary temporal complexity, but remains spatially
homogeneous by construction. The linear solution family (\ref{eq: Linear velocity field})
identified by Craik and Criminale \cite{craik1986evolution} cannot,
therefore, produce bounded coherent flow structures. As a consequence,
these linear solutions cannot yield Navier--Stokes flows with finite
coherent vortices or bounded chaotic mixing zones.

Higher-order polynomial vector fields of the form (\ref{eq: Full velocity field, general form}),
however, are free from these limitations, providing an endless source
of unsteady and dynamically consistent examples of flow structures
away from boundaries. We will construct specific examples of such
flows and illustrate how the instantaneous streamlines, as well as
the 2D version of the broadly used $Q$-criterion of Hunt, Wray
and Moin \cite{hunt1988eddies}, the $\Delta$-criterion of \cite{chong1990general},
the $\lambda_{2}$-criterion of \cite{jeong1995identification} and
the $\lambda_{ci}$-criterion (or swirling-strength criterion) of
\cite{chakraborty2005relationships}, fail in describing fluid particle
behavior correctly in these examples. We will also point out how the
two-dimensional exact solutions we construct explicitly extend to
unsteady 3D Navier--Stokes solutions.

\section{Universal Navier--Stokes solutions }

We rewrite the incompressible Navier--Stokes equation (under potential
body forces) for a velocity field $\mathbf{u}(\mathbf{x},t)$ in the
form 
\begin{equation}
\frac{\partial\mathbf{u}}{\partial t}+(\mathbf{u}\cdot\boldsymbol{\nabla})\mathbf{u}-\nu\text{\ensuremath{\Delta}}\mathbf{u}=\text{\ensuremath{\nabla}}\left[\text{\textminus}\frac{p+V}{\rho}\right],\label{eq: NS equation}
\end{equation}
where $\nu\geq0$ is the kinematic viscosity, $\rho>0$ is the density,
$p\left(\mathbf{x},t\right)$ is the pressure field and $V(\mathbf{x},t)$
is the potential of external body forces, such as gravity. We note
that the left-hand side of (\ref{eq: NS equation}) is a gradient
(i.e., conservative) vector field. By classic results in potential
theory, a vector field is conservative on a simply connected domain
if and only if its curl is zero. For a 2D vector field, this zero-curl
condition is equivalent to the requirement that the Jacobian of the
vector field is zero, as already noted in the construction of linear
Navier--Stokes solutions by Craik and Criminale \cite{craik1986evolution}.
On open, simply connected domains, therefore, a sufficient and necessary
condition for $\mathbf{u}(\mathbf{x},t)$ to be a Navier--Stokes
solution is given by 
\begin{widetext}
\begin{equation}
\boldsymbol{\nabla}\left[\frac{\partial\mathbf{u}}{\partial t}+(\mathbf{u}\cdot\boldsymbol{\nabla})\mathbf{u}-\nu\text{\ensuremath{\Delta}}\mathbf{u}\right]=\left(\boldsymbol{\nabla}\left[\frac{\partial\mathbf{u}}{\partial t}+(\mathbf{u}\cdot\boldsymbol{\nabla})\mathbf{u}-\nu\text{\ensuremath{\Delta}}\mathbf{u}\right]\right)^{T},\label{eq: Symmetry condition}
\end{equation}
\end{widetext}
which no longer depends on the pressure and the external body force
potential $V$. Substituting (\ref{eq: Full velocity field, general form})
into (\ref{eq: Symmetry condition}) and equating equal powers of
$x$ and $y$ in the off-diagonal elements of the matrices on the
opposite sides of the resulting equation, we obtain conditions on
the unknown coefficients of the spatially polynomial velocity field
$\mathbf{u}(\mathbf{x},t)$.

By a \emph{universal solution} of equation (\ref{eq: NS equation})
we mean a solution $\mathbf{u}(\mathbf{x},t)$ on which viscous forces
identically vanish, rendering the pressure $p(\boldsymbol{x},t)$
independent of the Reynolds number. Note that all spatially linear
solutions of (\ref{eq: NS equation}) are universal. More generally,
a solution $\mathbf{u}(\mathbf{x},t)$ of equation (\ref{eq: NS equation})
is a universal solution of the planar Navier--Stokes equation if
and only if 
\begin{equation}
\text{\ensuremath{\Delta}}\mathbf{u}\text{\ensuremath{\equiv}}\boldsymbol{0},\label{eq: Universal solution condition}
\end{equation}
i.e., if it is a harmonic solution. Hence, when looking for universal
solutions of the form (\ref{eq: Full velocity field, general form})
that satisfy (\ref{eq: Symmetry condition}), we look for solutions
$\mathbf{u}(\mathbf{x},t)$ whose components are harmonic polynomials
in $\mathbf{x}$ with time-dependent coefficients. Since the viscous
terms in the Navier--Stokes equation vanish for harmonic flows, the
universal solutions we find will also be solutions of Euler's equation.
Our main result is as follows: 
\begin{thm}
An $n^{th}$-order, unsteady polynomial velocity field $\mathbf{u}(\mathbf{x},t)=(u(x,y,t),v(x,y,t))$
of the spatial variable $\mathbf{x}=(x,y)$ is a universal solution
of the planar, incompressible Navier--Stokes equation (\ref{eq: NS equation})
if and only if 
\begin{widetext}
\begin{equation}
\mathbf{u}(\mathbf{x},t)=\mathbf{h}(t)+\frac{1}{2}\omega\left(\begin{array}{r}
-y\\
x
\end{array}\right)+\sum_{k=1}^{n}\left(\begin{array}{cr}
a_{k}(t) & b_{k}(t)\\
b_{k}(t) & -a_{k}(t)
\end{array}\right)\left(\begin{array}{c}
\mathrm{Re}\left(x+iy\right)^{k}\\
\mathrm{Im}\left(x+iy\right)^{k}
\end{array}\right)\label{eq: Exact solution form}
\end{equation}
\end{widetext}
holds for some arbitrary smooth functions $\mathbf{h}:\mathbb{R}\to\mathbb{R}^{2}$,
$a_{k},b_{k}\colon\mathbb{R}\to\mathbb{R}$ and an arbitrary constant
$\omega\in\mathbb{R}$, where $\omega$ coincides with the constant
scalar vorticity field of $\mathbf{u}(\mathbf{x},t)$. 
\end{thm}

\emph{Proof:} To prove that a universal solution $\mathbf{u}(\mathbf{x},t)$
must be precisely of the form given in (\ref{eq: Exact solution form}),
we recall from Andrews, Askey and Roy \cite{andrews2000special}
that a basis of the space of $k^{th}$-order, homogeneous harmonic
polynomials of two variables is given by 
\begin{equation}
\begin{aligned}\begin{split}&\left\{ \mathrm{Re}\left[f_{k}(x,y)\right],\mathrm{Im}\left[f_{k}(x,y)\right]\right\} ,\end{split}
\\
f_{k}(x,y):=\left(x+iy\right)^{k},\quad k=0,1,2,\ldots
\end{aligned}
\end{equation}
Hence the most general form of a polynomial $\mathbf{u}(\mathbf{x},t)$
satisfying the universality condition \eqref{eq: Universal solution condition}
is
\begin{equation}
\mathbf{u}(\mathbf{x},t)=\sum_{k=0}^{n}\left(\begin{array}{cc}
\alpha_{k}(t) & \beta_{k}(t)\\
\gamma_{k}(t) & \delta_{k}(t)
\end{array}\right)\left(\begin{array}{c}
\mathrm{Re}\left[f_{k}(x,y)\right]\\
\mathrm{Im}\left[f_{k}(x,y)\right]
\end{array}\right)\label{eq: General short solution form}
\end{equation}
for some smooth scalar-valued functions $\alpha_{k}(t),\beta_{k}(t),\gamma_{k}(t)$
and $\delta_{k}(t)$. By the Cauchy-Riemann equations for holomorphic
complex functions, $f_{k}(x,y)$ must then satisfy
\begin{equation}
\frac{\partial\mathrm{Re}f_{k}}{\partial x}=\frac{\partial\mathrm{Im}f_{k}}{\partial y},\label{eq: CR 1}
\end{equation}
\begin{equation}
\frac{\partial\mathrm{Re}f_{k}}{\partial y}=-\frac{\partial\mathrm{Im}f_{k}}{\partial x}.\label{eq: CR 2}
\end{equation}
By these conditions, requiring the divergence of $\mathbf{u}(\mathbf{x},t)$
in (\ref{eq: General short solution form}) to vanish is equivalent
to
\begin{equation}
\begin{split}\sum_{k=0}^{n}\left(\alpha_{k}\frac{\partial\mathrm{Re}f_{k}}{\partial x}+\beta_{k}\frac{\partial\mathrm{Im}f_{k}}{\partial x}+\gamma_{k}\frac{\partial\mathrm{Re}f_{k}}{\partial y}+\delta_{k}\frac{\partial\mathrm{Im}f_{k}}{\partial y}\right) & =\\
\sum_{k=0}^{n}\left(\left(\alpha_{k}+\delta_{k}\right)\frac{\partial\mathrm{Re}f_{k}}{\partial x}+\left(\gamma_{k}-\beta_{k}\right)\frac{\partial\mathrm{Re}f_{k}}{\partial y}\right) & \equiv0.
\end{split}
\label{eq: Divergence vanishing condition}
\end{equation}
For formula (\ref{eq: Divergence vanishing condition}) to hold at
order $k=0$, any constant term 
\begin{equation}
\mathbf{h}(t):=\left(\begin{array}{cc}
\alpha_{0}(t) & \beta_{0}(t)\\
\gamma_{0}(t) & \delta_{0}(t)
\end{array}\right)\left(\begin{array}{c}
1\\
1
\end{array}\right)\label{eq:k=00003D0}
\end{equation}
can be selected. At order $k=1$, the same formula requires
\[
\alpha_{1}(t)\equiv-\delta_{1}(t).
\]
Finally, for $k\ge2$, formula (\ref{eq: Divergence vanishing condition})
requires 
\[
\alpha_{k}(t)\equiv-\delta_{k}(t),\quad\beta_{k}(t)\equiv\gamma_{k}(t),\quad k\geq2.
\]

By equations (\ref{eq: CR 1})-(\ref{eq: CR 2}), the vorticity field
of the 2D velocity field $\mathbf{u}(\mathbf{x},t)$ is
\begin{widetext}
\[
\begin{split}\left[\boldsymbol{\nabla}\times\mathbf{u}(\mathbf{x},t)\right]_{3}=\frac{\partial v}{\partial x}-\frac{\partial u}{\partial y}= & \sum_{k=0}^{n}\left(\gamma_{k}\frac{\partial\mathrm{Re}f_{k}}{\partial x}-\alpha_{k}\frac{\partial\mathrm{Im}f_{k}}{\partial x}\right)-\left(\alpha_{k}\frac{\partial\mathrm{Re}f_{k}}{\partial y}+\beta_{k}\frac{\partial\mathrm{Im}f_{k}}{\partial y}\right) \\
= & \sum_{k=0}^{n}\left(\gamma_{k}\frac{\partial\mathrm{Re}f_{k}}{\partial x}-\beta_k \frac{\partial\mathrm{Im}f_{k}}{\partial y}\right)-\alpha_{k}\left(\frac{\partial\mathrm{Im}f_{k}}{\partial x}+\frac{\partial\mathrm{Re}f_{k}}{\partial y}\right)\\
= & \gamma_{1}(t)-\beta_{1}(t).
\end{split}
\]
\end{widetext}
Therefore, with the notation
\begin{align}
a_{k}(t) & :=\alpha_{k}(t)=-\delta_{k}(t),\quad k\geq1,\nonumber \\
b_{1}(t) & :=\frac{1}{2}\left(\beta_{1}(t)+\gamma_{1}(t)\right),\quad\omega(t):=\gamma_{1}(t)-\beta_{1}(t),\label{eq:k=00003D1,2}\\
b_{k}(t) & :=\beta_{k}(t)\equiv\gamma_{k}(t),\quad k\geq2,\nonumber 
\end{align}
and with the identity
\begin{widetext}
\begin{equation*}
\begin{aligned}\left(\begin{array}{cc}
\alpha_{1}(t) & \beta_{1}(t)\\
\gamma_{1}(t) & -\alpha_{1}(t)
\end{array}\right) & \left(\begin{array}{c}
\mathrm{Re}f_{1}\\
\mathrm{Im}f_{1}
\end{array}\right)=\\
\left(\begin{array}{cc}
\alpha_{1}(t) & \frac{1}{2}\left(\gamma_{1}(t)+\beta_{1}(t)\right)\\
\frac{1}{2}\left(\gamma_{1}(t)+\beta_{1}(t)\right) & -\alpha_{1}(t)
\end{array}\right) & \left(\begin{array}{c}
\mathrm{Re}f_{1}\\
\mathrm{Im}f_{1}
\end{array}\right)+\\
\left(\begin{array}{cc}
0 & \frac{1}{2}\left(\gamma_{1}(t)+\beta_{1}(t)\right)\\
\frac{1}{2}\left(\gamma_{1}(t)+\beta_{1}(t)\right) & 0
\end{array}\right) & \left(\begin{array}{c}
\mathrm{Re}f_{1}\\
\mathrm{Im}f_{1}
\end{array}\right)\\
=\left(\begin{array}{cc}
\alpha_{1}(t) & b_{1}(t))\\
b_{1}(t)) & -\alpha_{1}(t)
\end{array}\right)\left(\begin{array}{c}
\mathrm{Re}f_{1}\\
\mathrm{Im}f_{1}
\end{array}\right) & +\left(\begin{array}{cc}
0 & -\frac{1}{2}\omega(t)\\
\frac{1}{2}\omega(t) & 0
\end{array}\right)\left(\begin{array}{c}
x\\
y
\end{array}\right),
\end{aligned}
\end{equation*}
\end{widetext}
formulas \eqref{eq: General short solution form}-\eqref{eq:k=00003D1,2}
give 
\begin{widetext}
\begin{equation}
\mathbf{u}(\mathbf{x},t)=\mathbf{h}(t)+\frac{1}{2}\omega(t)\left(\begin{array}{r}
-y\\
x
\end{array}\right)+\sum_{k=1}^{n}\left(\begin{array}{cr}
a_{k}(t) & b_{k}(t)\\
b_{k}(t) & -a_{k}(t)
\end{array}\right)\left(\begin{array}{c}
\mathrm{Re}f_{k}\\
\mathrm{Im}f_{k}
\end{array}\right),\label{eq: Exact solution form-1}
\end{equation}
\end{widetext}
with $\omega(t)$ denoting the spatially constant (but at this point
time-dependent) scalar vorticity field of \eqref{eq: Exact solution form-1}.
Since $\mathbf{u}(\mathbf{x},t)$ is harmonic and incompressible,
the symmetry condition (\ref{eq: Symmetry condition}) is satisfied
by the velocity field \eqref{eq: Exact solution form-1} if and only
if the $2\times2$ matrix 
\begin{equation}
\text{\ensuremath{\boldsymbol{\ensuremath{\nabla}}}}\frac{\partial\mathbf{u}}{\partial t}+\boldsymbol{\nabla}\left[(\mathbf{u}\cdot\boldsymbol{\nabla})\mathbf{u}\right]\label{eq:symmetric matrix}
\end{equation}
is symmetric. To verify the symmetry of this matrix using the notation
\textbf{$\mathbf{u}=\left(u,v\right)$}, first note that the skew-symmetric
parts of the two summands in \eqref{eq:symmetric matrix} are given
by
\begin{widetext}
\begin{align}
\mathrm{skew}\left[\text{\ensuremath{\boldsymbol{\ensuremath{\nabla}}}}\frac{\partial\mathbf{u}}{\partial t}\right] & =\frac{1}{2}\left(\boldsymbol{\nabla}\frac{\partial\boldsymbol{u}}{\partial t}-\left[\boldsymbol{\nabla}\frac{\partial\boldsymbol{u}}{\partial t}\right]^{T}\right)=\frac{1}{2}\left(\begin{array}{cr}
0 & -\dot{\omega}(t)\\
\dot{\omega}(t) & 0
\end{array}\right),\label{eq:skew1}\\
\mathrm{skew}\left[\boldsymbol{\nabla}\left[(\mathbf{u}\cdot\boldsymbol{\nabla})\mathbf{u}\right]\right] & =\mathrm{skew}\left(\begin{array}{cr}
0 & u_{xy}u+u_{yy}v+\left(u_{x}+v_{y}\right)u_{y}\\
v_{xx}u+v_{xy}v+\left(u_{x}+v_{y}\right)v_{x} & 0
\end{array}\right)=\mathbf{0},\label{eq:skew2}
\end{align}
\end{widetext}
where we have used the incompressibility condition $u_{x}+v_{y}=0,$
as well as the spatial independence of the scalar vorticity $\omega(t)$
of $\mathbf{u}$, which implies $\partial_{x}\omega=v_{xx}-u_{xy}\equiv0$
and $\partial_{y}\omega=v_{xy}-u_{yy}\equiv0$. Formulas \eqref{eq:symmetric matrix}-\eqref{eq:skew2}
then imply that for the matrix \eqref{eq:symmetric matrix} to be
symmetric, we must have $\dot{\omega}(t)\equiv0$, i.e., $\omega(t)=\omega=const.$
must hold in \eqref{eq: Exact solution form-1}. This completes the
proof of the theorem.~~~~~$\square$

For the universal solutions derived in this section, the pressure
field $p(\mathbf{x},t)$ can be obtained by substituting the solutions
into (\ref{eq: NS equation}), integrating the left-hand side of the
resulting equation, multiplying the result by $-\rho$ and subtracting
the potential $V(\mathbf{x},t)$. 

The two-dimensional Navier--Stokes solution family \eqref{eq: Exact solution form}
immediately generates three-dimensional, incompressible Navier--Stokes
solutions as well. The planar components of these solutions are just
given by \eqref{eq: Exact solution form}, while their vertical component,
$w(x,y,t)$, satisfies the scalar advection-diffusion equation 
\begin{equation}
\partial_{t}w+\boldsymbol{\nabla}w\cdot\mathbf{u}=\nu\Delta w,\label{eq:equation for w}
\end{equation}
 as shown, e.g., by Majda and Bertozzi \cite{majda2002vorticity}.
We therefore obtain the following result.
\begin{prop}
Any polynomial solution \eqref{eq: Exact solution form-1} of the
planar Navier--Stokes equation generates a family of exact solutions
$\left(\mathbf{u}(\mathbf{x},t),w(\mathbf{x},t)\right)$ for the three-dimensional
version of the Navier--Stokes equation \eqref{eq: NS equation},
where $w(\mathbf{x},t)$ is an arbitrary solution of the advection-diffusion
eq. \eqref{eq:equation for w}. In particular, \textbf{
\begin{widetext}
\begin{equation}
\mathbf{v}(\mathbf{x},t)=\left(\begin{array}{r}
\mathbf{u}(\mathbf{x},t)\\
w(\mathbf{x},t)
\end{array}\right)=\left(\begin{array}{c}
\mathbf{h}(t)\\
w_{0}
\end{array}\right)+\frac{1}{2}\omega(t)\left(\begin{array}{c}
\begin{array}{r}
-y\\
x
\end{array}\\
0
\end{array}\right)+\left(\begin{array}{c}
\sum_{k=1}^{n}\left(\begin{array}{cr}
a_{k}(t) & b_{k}(t)\\
b_{k}(t) & -a_{k}(t)
\end{array}\right)\left(\begin{array}{c}
\mathrm{Re}f_{k}\\
\mathrm{Im}f_{k}
\end{array}\right)\\
0
\end{array}\right),\label{eq:3D extension}
\end{equation}
\end{widetext}}is an exact, unsteady polynomial solution of the 3D Navier--Stokes
equation for any choice of the constant vertical velocity $w_{0}\in\mathbb{R}$. 
\end{prop}

\section{\label{sec:level2}vortex identification methods}

For our later analysis of universal solution examples, we now briefly
review the most broadly used pointwise structure identification schemes
in unsteady flows. For 2D flows, the Okubo--Weiss criterion (Okubo
\cite{okubo1970horizontal}, Weiss \cite{weiss1991dynamics}) postulates
that the nature of fluid particle motion in a flow is governed by
the eigenvalue configuration of the velocity gradient $\boldsymbol{\nabla}\mathbf{u}$$(\mathbf{x},t)$.
For 2D incompressible flows, this eigenvalue configuration is uniquely
characterized by the scalar field 

\begin{equation}
OW(\mathbf{x},t)=-\det\left[\boldsymbol{\nabla}\mathbf{u}(\mathbf{x},t)\right].\label{eq: OW criterion}
\end{equation}
The Okubo--Weiss criterion postulates that in elliptic (or vortical)
regions, the velocity gradient $\boldsymbol{\nabla}\mathbf{u}$$(\mathbf{x},t)$
has purely imaginary eigenvalues, or equivalently, $OW(\mathbf{x},t)<0$
holds. Similarly, the criterion postulates that hyperbolic (or stretching)
regions are characterized by $OW(\mathbf{x},t)>0$. 

As formula \eqref{eq:3D extension} shows, any universal 2D Navier--Stokes
solutions generates a 3D Navier--Stokes solution family $\mathbf{v}=\left(\mathbf{u},w_{0}\right)$
with an arbitrary, constant vertical velocity component $w_{0}$.
This extension enables the application of 3D local vortex-identification
criteria to our universal solutions. Such criteria generally involve
the spin tensor $\boldsymbol{\Omega}=\frac{1}{2}\left[\boldsymbol{\nabla}\mathbf{v}-\left(\boldsymbol{\nabla}\mathbf{v}\right)^{T}\right]$
and the rate-of-strain tensor $\mathbf{S}=\frac{1}{2}\left[\boldsymbol{\nabla}\mathbf{v}+\left(\boldsymbol{\nabla}\mathbf{v}\right)^{T}\right]$,
the antisymmetric and symmetric parts of $\nabla\mathbf{v}$, respectively. 

While Epps \cite{Epps2017} provides a comprehensive review of such
criteria, we will here focus on the three most broadly used ones.

We also mention the work of \citet{rousseaux2007lamb}, who utilize the hydrodynamic Lamb vector (cf. \citet{Belevich_2008}, \citet{doi:10.1063/1.869762} and \citet{PhysRevE.58.522}) in vortex detection. 

The first of the three local vortex criteria we discuss here, the $Q$-criterion of Hunt, Wray and Moin \cite{hunt1988eddies},
postulates that in a vortical region, the Euclidean matrix norm of
$\boldsymbol{\Omega}$ dominates that of $\boldsymbol{\mathbf{S}}$,
rendering the scalar field
\begin{equation}
Q(\mathbf{x},t)=\frac{1}{2}\left(\left\Vert \boldsymbol{\Omega}\right\Vert ^{2}-\left\Vert \boldsymbol{\mathbf{S}}\right\Vert ^{2}\right)\label{eq:Q criterion}
\end{equation}
positive. One can verify by direct calculation that $Q(\mathbf{x},t)\equiv OW(\mathbf{x},t)$
holds for the 3D extension \eqref{eq:3D extension} of our universal
solutions (and, in general, for any two-dimensional flow), rendering
the $Q$-criterion equivalent to the Okubo--Weiss criterion for these
flows.

Second, the $\Delta$-criterion of Chong et al. \cite{chong1990general}
seeks vortices in 3D flow as domains where the velocity gradient $\boldsymbol{\nabla}\mathbf{v}(\mathbf{x},t)$
admits eigenvalues with nonzero imaginary parts. In the extended universal
solutions \eqref{eq:3D extension}, zero is always an eigenvalue for
$\boldsymbol{\nabla}\mathbf{v}$ and hence, by incompressibility,
the remaining two eigenvalues of $\boldsymbol{\nabla}\mathbf{v}$
are complex precisely when they are purely imaginary. This happens
precisely when $Q(\mathbf{x},t)\equiv OW(\mathbf{x},t)>0$ holds,
and hence the $\Delta$-criterion also coincides with the Okubo--Weiss
criterion on the universal solutions we have constructed. 

Third, the $\lambda_{2}$-criterion of Jeong and Hussein \cite{jeong1995identification}
identifies vortices as the collection of points where the pressure
has a local minimum within an appropriately chosen two-dimensional
plane. Under various further assumptions, this principle is equivalent
to the requirement that the intermediate eigenvalue, $\lambda_{2}\left(\mathbf{S}^{2}+\mathbf{W}^{2}\right)$,
of the symmetric tensor $\mathbf{S}^{2}+\mathbf{W}^{2}$ must satisfy.
\begin{equation}
\lambda_{2}\left(\mathbf{S}^{2}+\mathbf{W}^{2}\right)<0.\label{eq:lambda_2 criterion}
\end{equation}
For the extended universal solutions \eqref{eq:3D extension}, we
obtain from a direct calculation that
\begin{align*}
\lambda_{2}\left(\mathbf{S}^{2}+\mathbf{W}^{2}\right) & =S_{11}^{2}+S_{12}^{2}-W_{12}^{2}=-Q,
\end{align*}
the $\lambda_{2}$-criterion also agrees with Okubo--Weiss criterion
for these flows. 

Finally, the $\lambda_{ci}$-criterion (or swirling-strength criterion)
of Chakraborthy et al. \cite{chakraborty2005relationships} follows
the logic of the $\Delta$-criterion and asserts that local material
swirling occurs at points where $\boldsymbol{\nabla}\mathbf{v}$ has
a pair of complex eigenvalues $\lambda_{cr}+i\lambda_{ci}$ and a
real eigenvalue $\lambda_{r}$. To ensure tight enough spiraling (orbital
compactness) typical for a vortex, the $\lambda_{ci}$-criterion asserts
the requirement
\begin{equation}
\lambda_{ci}\geq\epsilon,\qquad\lambda_{cr}/\lambda_{ci}\leq\delta,\label{eq:lambda_ci}
\end{equation}
for some small, constant thresholds $\epsilon,\delta>0.$ Again, for
the extended universal flows \eqref{eq:3D extension}, we can only
have $\lambda_{ci}=0$ and on any compact domain with $\lambda_{ci}>0$
we can select an $\epsilon>0$ such that the criterion \eqref{eq:lambda_2 criterion}
is satisfies. Once again, therefore, the $\lambda_{ci}$-criterion
agrees with the Okubo--Weiss criterion for the the extended universal
flows \eqref{eq:3D extension}.

All these vortex criteria seek to capture vortex-type fluid particle
behavior in a heuristic fashion, i.e., hope to achieve conclusions
about the trajectories generated by the velocity field $\mathbf{v}\left(\mathbf{x},t\right)$
from instantaneous snapshots of $\mathbf{v}\left(\mathbf{x},t\right)$.
Some of these criteria have been shown to give reasonable results
on simple steady flows. Such steady examples have prompted the broad
use of these criteria in analyzing general unsteady flow data for
which no ground truth is available. This practice can immediately
be questioned based on first principles, given that all these criteria
are frame-dependent (see Haller \cite{haller2005objective,haller2015lagrangian}),
yet truly unsteady flows have no distinguished frames (cf. Lugt \cite{lugt1979dilemma}).
Another questionable practice has been to simply replace the original
versions of vortex identification criteria with plots of heuristically
chosen level surfaces of the quantities arising in these criteria.
Such level surfaces are generally highly sensitive to the choice of
the constant value that defines them, yet these constants are routinely
chosen to match intuition or provide visually pleasing results.

The exact, unsteady Navier--Stokes solutions we have constructed
here give an opportunity to test the vortex identification
methods above on nontrivial, unsteady Navier--Stokes solutions in which
a ground truth can be reliably established via Poincaré maps. As we
will see below, such a systematic analysis reveals major inconsistencies
for all the vortex criteria recalled above. As we have seen, it will
be enough to point out these inconsistencies on the 2D universal solutions
\eqref{eq: Exact solution form} for the Okubo--Weiss criterion.
The same inconsistencies arise then automatically in the analysis
of the extended 3D Navier--Stokes solution \eqref{eq:3D extension}
via the $Q$-, $\Delta$-, $\lambda_{2}$- and $\lambda_{ci}$-criteria.

\section{Examples}

We now give examples of dynamically consistent flow fields covered
by the general formula (\ref{eq: Exact solution form}). With these
exact solutions, we illustrate that using the instantaneous streamlines
for material vortex detection (see, e.g., Sadajoen and Post \cite{sadarjoen2000detection}
and Robinson \cite{robinson1991coherent}) gives inconsistent results.
With the same solutions, we also illustrate how the Okubo--Weiss
criterion described in the previous section fails to identify the
true nature of unsteady fluid particle motion. 
\begin{example}
\label{Example 1}Haller \cite{haller2015lagrangian,haller2005objective}
proposed the velocity field 
\begin{equation}
\mathbf{u}(\mathbf{x},t)=\left(\begin{array}{cc}
\sin4t & \cos4t+2\\
\cos4t-2 & -\sin4t
\end{array}\right)\mathbf{x},\label{eq: Linear example 1}
\end{equation}
as a purely kinematic benchmark example for testing vortex criteria.
By inspection of (\ref{eq: Exact solution form}), we find that (\ref{eq: Linear example 1})
solves the Navier--Stokes equation with $\mathbf{h}(t)\equiv\mathbf{0}$,
$a_{1}(t)=\sin4t$, $b_{1}(t)=\cos4t$, $\omega=-4$, and $a_{k}=b_{k}\equiv0$
for $k\geq2$. More generally, formula (\ref{eq: Exact solution form})
shows that the linear unsteady velocity field 
\begin{equation}
\dot{\mathbf{x}}=\mathbf{u}(\mathbf{x},t)=\left(\begin{array}{cc}
-\sin Ct & \cos Ct-\frac{\omega}{2}\\
\cos Ct+\frac{\omega}{2} & \sin Ct
\end{array}\right)\mathbf{x}\label{eq:Generalized linear example}
\end{equation}
solves the 2D Navier--Stokes equation for any constants $\omega$
and $C$, and any smooth function $\mathbf{h}(t)$. We set $\mathbf{h}(t)\equiv\mathbf{0}$
for simplicity and pass to a rotating $\mathbf{y}$-coordinate frame
via the transformation 
\begin{equation}
\mathbf{x}=\mathbf{M}(t)\mathbf{y},\quad\mathbf{M}(t)=\left(\begin{array}{cc}
\cos\frac{C}{2}t & \sin\frac{C}{2}t\\
-\sin\frac{C}{2}t & \cos\frac{C}{2}t
\end{array}\right).\label{eq: Frame change}
\end{equation}

In these new coordinates, (\ref{eq:Generalized linear example}) becomes
\begin{align}
\dot{\mathbf{x}} & =\left(\begin{array}{cc}
-\sin Ct & \cos Ct-\frac{\omega}{2}\\
\cos Ct+\frac{\omega}{2} & \sin Ct
\end{array}\right)\mathbf{x}\nonumber \\
 & =\left(\begin{array}{cc}
-\sin Ct & \cos Ct-\frac{\omega}{2}\\
\cos Ct+\frac{\omega}{2} & \sin Ct
\end{array}\right)\mathbf{M}^{-1}\mathbf{y}\\
 & =\left(\begin{array}{cc}
-\left(1+\frac{\omega}{2}\right)\sin\frac{C}{2}t & \left(1-\frac{\omega}{2}\right)\cos\frac{C}{2}t\\
\left(1+\frac{\omega}{2}\right)\cos\frac{C}{2}t & \left(1-\frac{\omega}{2}\right)\sin\frac{C}{2}t
\end{array}\right)\mathbf{y}.\label{eq:x dot in new frame}
\end{align}
At the same time, differentiating the coordinate change (\ref{eq: Frame change})
with respect to time gives $\dot{\mathbf{x}}=\dot{\mathbf{M}}\mathbf{y+}\mathbf{M}\dot{\mathbf{y}}$,
which, combined with \eqref{eq:x dot in new frame} gives the velocity
field in the $\mathbf{y}$-frame as 
\begin{equation}
\dot{\mathbf{y}}=\tilde{\mathbf{u}}(\mathbf{y})=\left(\begin{array}{cc}
0 & 1+\frac{1}{2}\left(C-\omega\right)\\
1-\frac{1}{2}\left(C-\omega\right) & 0
\end{array}\right)\mathbf{y}.\label{eq: Linear velocity field in rotating frame}
\end{equation}
This transformed velocity field is steady, defining an exactly solvable
autonomous linear system of differential equations for particle motions.
The nature of its solutions depends on the eigenvalues $\lambda_{1,2}=\pm\sqrt{1-\frac{1}{4}\left(\omega-C\right)^{2}}$
of the coefficient matrix in (\ref{eq: Frame change}). Specifically,
for $|\omega-C|<2$, we have a saddle-type flow with typical solutions
growing exponentially, while for $|\omega-C|>2$, we have a center-type
flow in which all trajectories perform periodic motion.

\begin{figure*}
\includegraphics[viewport=100bp 90bp 1900bp 1000bp,scale=0.28]{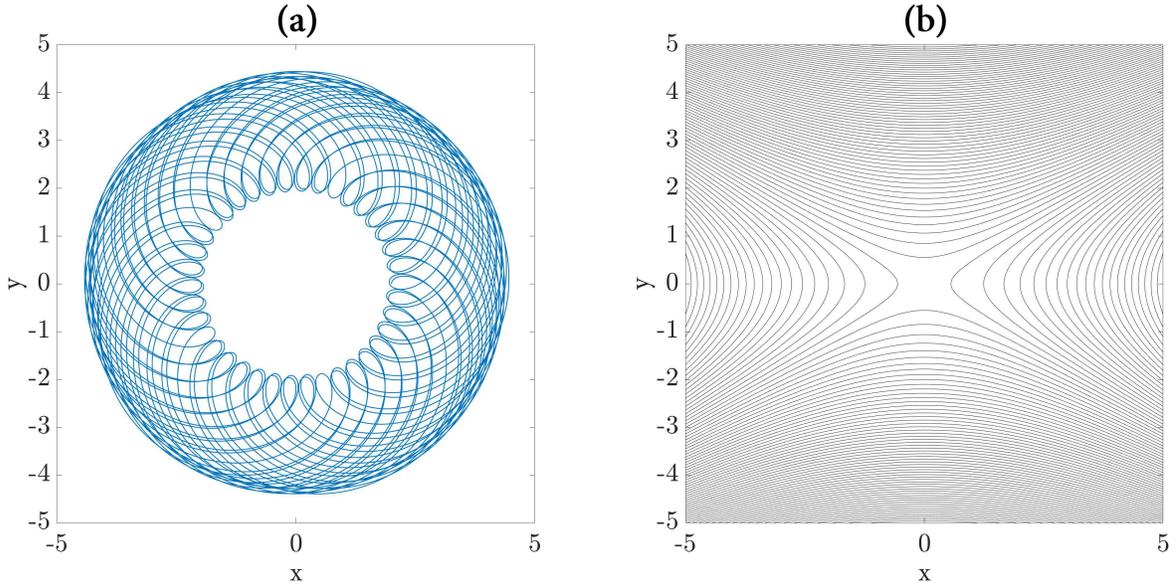}
\caption{(a) A typical fluid particle trajectory generated by the linear unsteady
velocity field (\ref{eq: Linear example 2}) for the time interval
$\left[t_{0},t_{1}\right]=\left[0,200\right]$ for the initial condition
$\boldsymbol{x}_{0}=(2,0)$. This flow would appear as an unbounded
vortex in any flow-visualization experiment involving dye or particles.
(b) The instantaneous streamlines of the same velocity field, shown
here for $t=0$, suggest a saddle point at the origin for all times
(streamlines at other times look similar).\label{fig: Linear example}}
\end{figure*}

Mapped back into the original frame via the time-periodic transformation
(\ref{eq: Frame change}), the center-type trajectories become quasiperiodic.
Figure \ref{fig: Linear example}(a) shows such a quasiperiodic particle
trajectory and Fig.~\ref{fig: Linear example}(b) instantaneous streamlines
of the velocity field (\ref{eq: Exact solution form}) with $\mathbf{h}(t)\equiv\mathbf{0}$,
$a_{1}(t)=\mathrm{sin}4t$, $b_{1}(t)=\mathrm{cos}4t$, and $\omega=-1$
and $a_{k}=b_{k}\equiv0$ for $k\ge2$. These parameter values yield
the velocity field 
\begin{equation}
\mathbf{u}(\mathbf{x},t)=\left(\begin{array}{cc}
\sin4t & \cos4t+\frac{1}{2}\\
\cos4t-\frac{1}{2} & -\sin4t
\end{array}\right)\mathbf{x},\label{eq: Linear example 2}
\end{equation}
which satisfies $|\omega-C|>2$. This flow would, therefore, appear
as an unbounded vortex in any flow visualization experiment involving
dye or particles, yet its instantaneous streamlines suggest a saddle
point at the origin for all times. Similarly, the Okubo--Weiss criterion
incorrectly pronounces the entire plane hyperbolic for the flow (\ref{eq: Linear example 2})
for all times. Indeed, formula (\ref{eq: OW criterion}) gives 
\begin{equation}
OW\equiv1-\frac{\omega^{2}}{4}=\frac{7}{16}>0.
\end{equation}
\end{example}

\begin{example}
\label{Example 2}By the general formula (\ref{eq: Exact solution form}),
a simple quadratic extension of the linear velocity field (\ref{eq: Linear example 1})
is given by the universal Navier--Stokes solution 
\begin{equation}
\mathbf{u}(\mathbf{x},t)=\left(\begin{array}{cc}
\sin4t & \cos4t+2\\
\cos4t-2 & -\sin4t
\end{array}\right)\mathbf{x}+\alpha(t)\left(\begin{array}{c}
x^{2}-y^{2}\\
-2xy
\end{array}\right),\label{eq:Quadratic example 1}
\end{equation}
where we have chosen $a_{2}(t)\equiv\alpha(t)$ and $b_{2}(t)\equiv0$
in the quadratic terms of (\ref{eq: Exact solution form}), and set
$\mathbf{h}(t)$, $a_{k}(t)$ and $b_{k}$ for $k>2$, as in Example
1. Setting $\alpha(t)\equiv-0.1$ for simplicity, we find that the
instantaneous streamlines now suggest a bounded spinning vortex enclosed
by connections between two stagnation points. The Okubo--Weiss criterion
also suggests a coherent vortex surrounding the origin at all times,
as $OW<0$ holds on a yellow domain shown containing the origin, as
shown in Fig.~\ref{fig: Quadratic example 1}(a). In reality, however,
the origin is a saddle-type Lagrangian trajectory with transversely
intersecting stable and unstable manifolds.

\begin{figure*}
\includegraphics[viewport=100bp 90bp 1900bp 1000bp,scale=0.28]{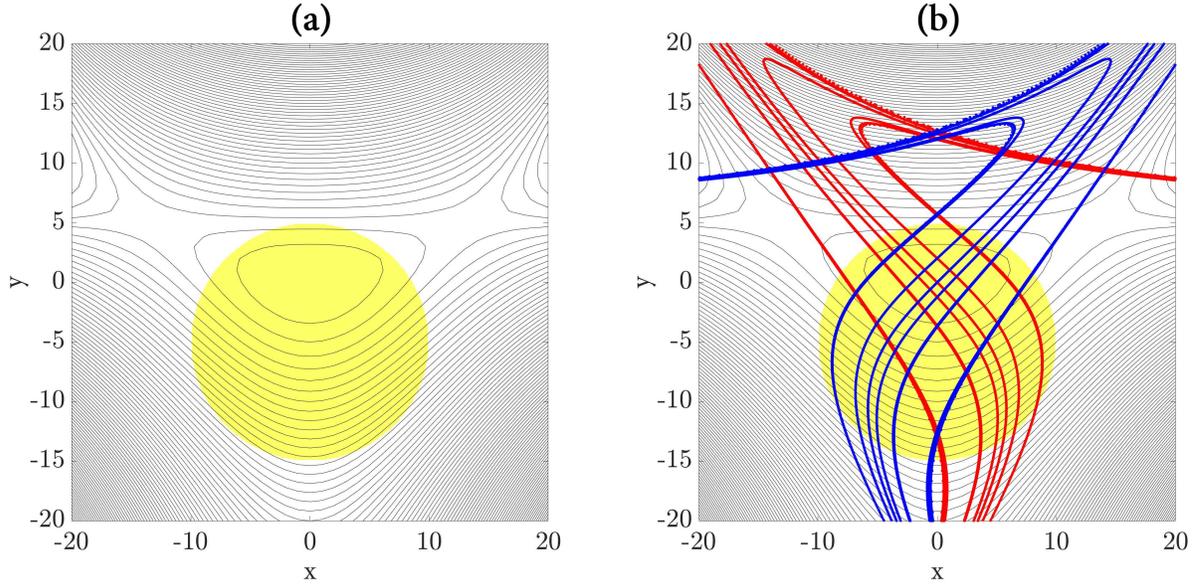}\caption{(a) Instantaneous streamlines and Okubo--Weiss elliptic region (yellow)
for the universal Navier--Stokes solution (\ref{eq:Quadratic example 1})
with $\alpha(t)\equiv-0.1$ at time $t=0$. Other time slices are
similar. (b) Stable (blue) and unstable (red) manifolds of the fixed
point of the Poincaré map (based at $t=0$ with period $T=\frac{\pi}{2}$)
for the Lagrangian particle motions under the same velocity field,
superimposed on the structures shown in (a).\label{fig: Quadratic example 1}}
\end{figure*}

Shown in Fig.~\ref{fig: Quadratic example 1}(b) for the Poincaré 
map of the flow, the resulting homoclinic tangle creates intense chaotic
mixing. This mixing process rapidly removes all but a measure zero
set of initial conditions from the Okubo--Weiss vortical region.
Therefore, the Navier--Stokes solution (\ref{eq:Quadratic example 1})
with $\alpha(t)\equiv-0.1$ provides a clear false positive for coherent
material vortex detection based on streamlines and on the Okubo--Weiss
criterion. 
\end{example}

\begin{example}
Building on the discussion of the stability of the $x=0$ fixed point
of equation (\ref{eq:Generalized linear example}), we now consider
another specific Navier--Stokes velocity field of the form 
\begin{equation}
\mathbf{u}(\mathbf{x},t)=\left(\begin{array}{cc}
\sin4t & \cos4t+\frac{1}{2}\\
\cos4t-\frac{1}{2} & -\sin4t
\end{array}\right)\mathbf{x}+\alpha(t)\left(\begin{array}{c}
x^{2}-y^{2}\\
-2xy
\end{array}\right),\label{eq:Quadratic example 2}
\end{equation}

\begin{figure*}
\includegraphics[viewport=100bp 90bp 1900bp 1000bp,scale=0.28]{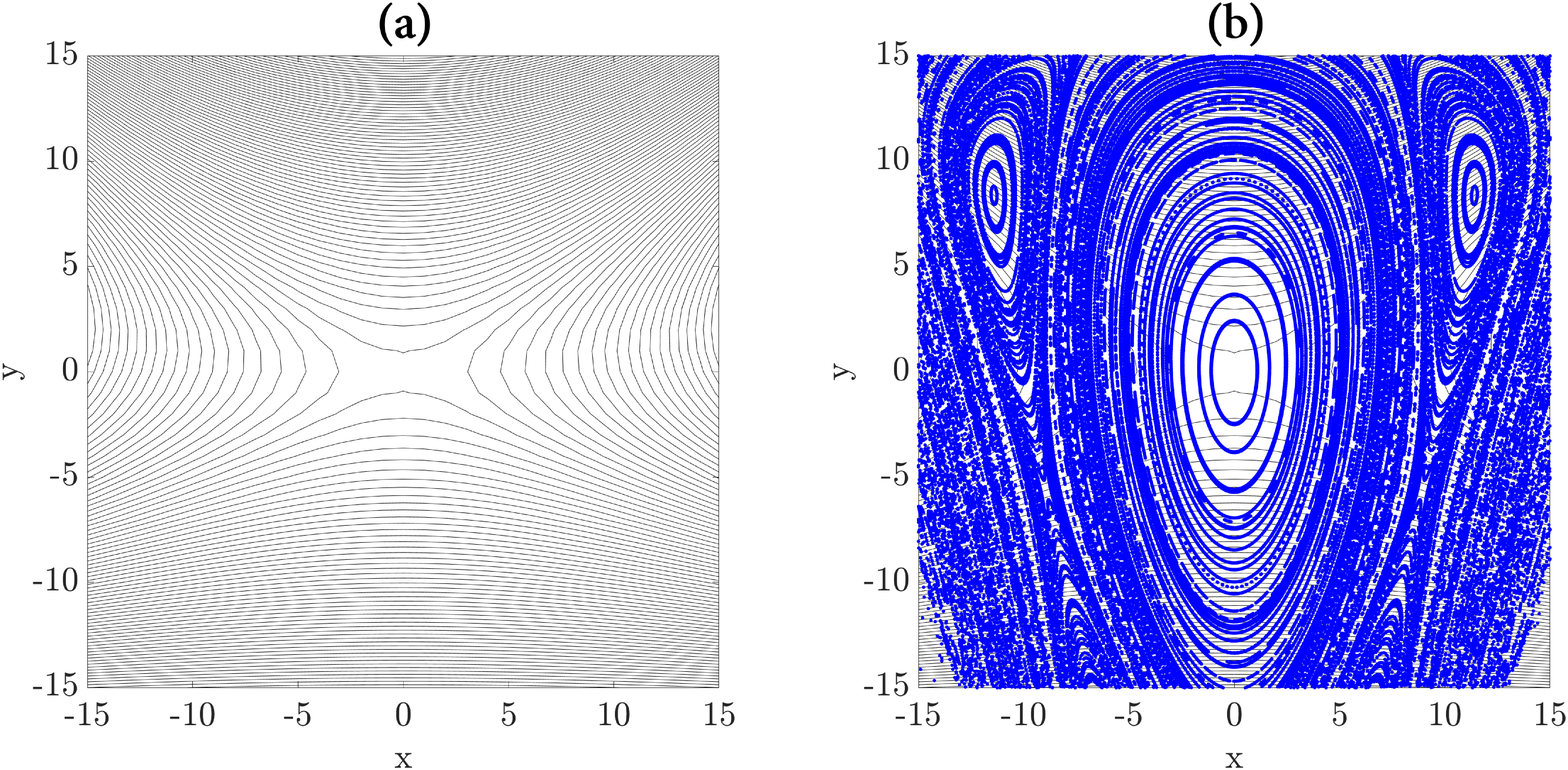}\caption{(a) Instantaneous streamlines for the universal Navier--Stokes solution
(\ref{eq:Quadratic example 2}) with $\alpha(t)\equiv-0.015$ at time
$t=0$. Other time slices are similar. (b) KAM curves (blue) of the
same velocity field, superimposed on the streamlines shown in (a).\label{fig: Quadratic example 2}}
\end{figure*}
from the universal solution family (\ref{eq: Exact solution form}).
In the notation used for equation (\ref{eq:Generalized linear example}),
we now have $\omega=-1$ and $C=4,$ which gives $|C-\omega|>2$.
Therefore, as discussed in Example 1, the origin of (\ref{eq:Quadratic example 2})
is a center-type fixed point under linearization for the Lagrangian
particle motion. At the same time, both the instantaneous streamlines
in Fig.~\ref{fig: Quadratic example 2}(a) and the Okubo--Weiss
criterion suggest saddle-type (hyperbolic) behavior for the linearized
flow, given that $OW>0$ holds on the whole plane. By the Kolmogorov--Arnold--Moser
(KAM) theorem \cite{arnold1989mathematical}, however, setting the
small parameter $\alpha(t)\equiv-0.015$ in (\ref{eq:Quadratic example 2})
is expected to preserve the elliptic (vortical) nature of the Lagrangian
particle motion in the quadratic velocity field (\ref{eq:Quadratic example 2}).
Indeed, most quasiperiodic motions of the linearized system survive
with the exception of resonance islands, as indicated by the KAM curves
shown in blue in Fig.~\ref{fig: Quadratic example 2}(b). Therefore,
the Navier--Stokes solution (\ref{eq:Quadratic example 2}) with
$\alpha(t)\equiv-0.015$ provides a false negative for coherent material
vortex detection based on streamlines or the Okubo--Weiss criterion.
Note that the KAM curves shown in some of the examples in this section
were obtained by launching fluid particles from a uniformly spaced
grid over the domain shown, advecting them over the time span $\left[0\ 2\pi\right]$,
and plotting the advected positions of the fluid particles at each
time step. 
\end{example}

\begin{example}
By the general formula (\ref{eq: Exact solution form}), a cubic extension
of (\ref{eq: Linear example 1}) is given by the universal Navier--Stokes
solution
\begin{widetext}
\begin{equation}
\mathbf{u}(\mathbf{x},t)=\left(\begin{array}{cc}
\sin4t & \cos4t+2\\
\cos4t-2 & -\sin4t
\end{array}\right)\mathbf{x}+\alpha(t)\left(\begin{array}{c}
x\left(x^{2}-3y^{2}\right)\\
-y\left(3x^{2}-y^{2}\right)
\end{array}\right),\label{eq: Cubic example}
\end{equation}
\end{widetext}
where we have chosen $a_{2}(t)\equiv0$, $b_{2}(t)\equiv0$, $a_{3}(t)\equiv\alpha(t)$
and $b_{3}(t)\equiv0$ in the quadratic and cubic terms of (\ref{eq: Exact solution form}),
respectively, and set $\mathbf{h}(t)$, $\omega$, $a_{k}(t)$ and
$b_{k}(t)$ for $k>3$ as in Example 2. We also set $\alpha(t)\equiv0.005$.
The instantaneous streamlines, shown in Fig.~\ref{fig: Cubic Example 1}(a)
for $t=0$, suggest a bounded spinning vortex around the origin surrounded
by two saddle-type structures. Similarly, the Okubo--Weiss criterion,
visualized by the yellow domain ($OW<0$) in Fig.~\ref{fig: Cubic Example 1}(a),
suggests a coherent vortex surrounding the origin at all times.

\begin{figure*}
\includegraphics[viewport=100bp 90bp 1900bp 1000bp,scale=0.28]{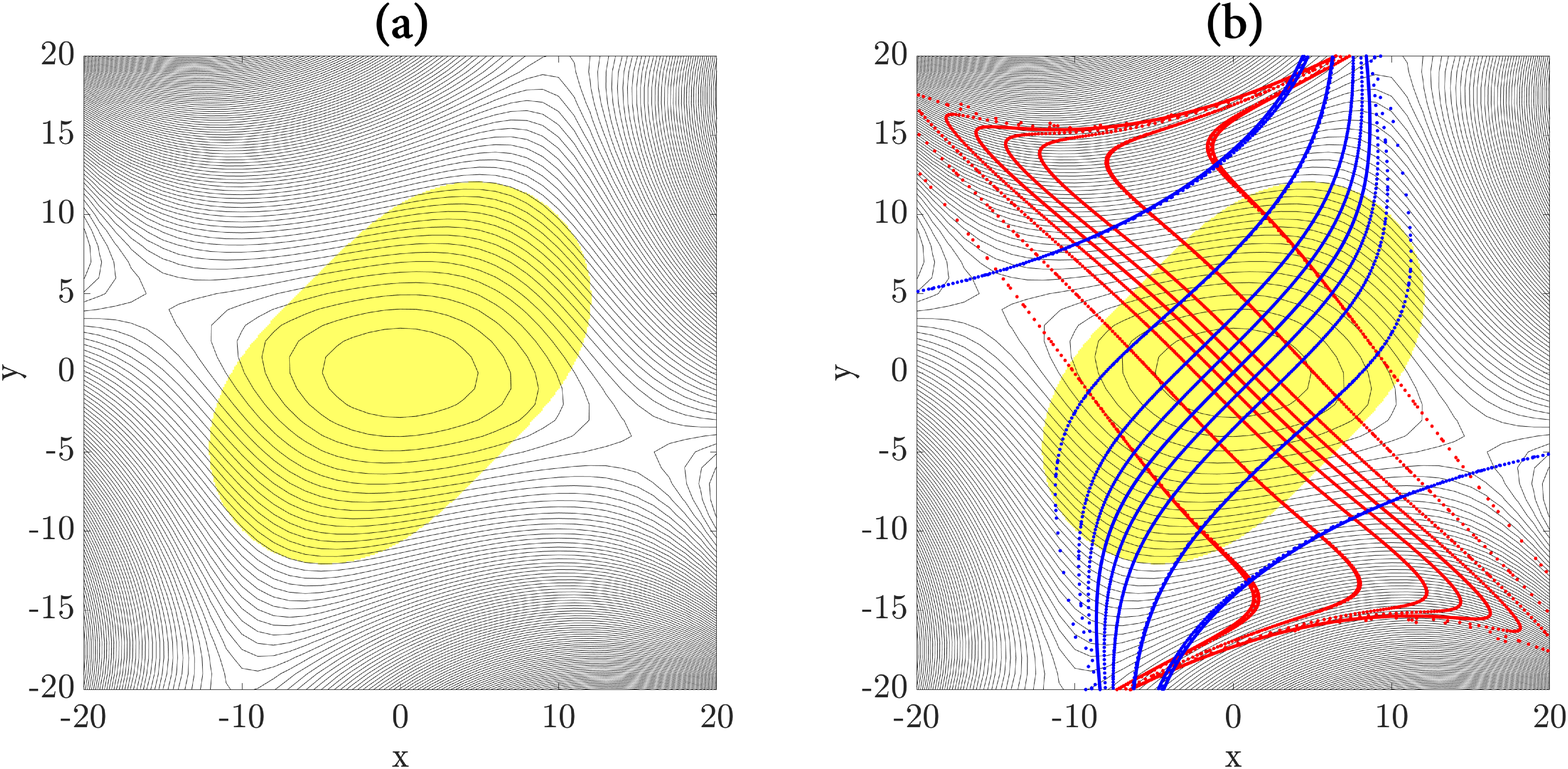}\caption{(a) Instantaneous streamlines and Okubo--Weiss elliptic region (yellow)
for the universal Navier--Stokes solution (\ref{eq: Cubic example})
with $\alpha(t)\equiv0.005$ at time $t=0$. Other time slices are
similar. (b) Stable (blue) and unstable (red) manifolds of the fixed
point of the Poincaré map (based at $t=0$ with period $T=\frac{\pi}{2}$)
for the Lagrangian particle motions under the same velocity field,
superimposed on the structures shown in (a).\label{fig: Cubic Example 1}}
\end{figure*}

The actual Lagrangian dynamics, however, is again strikingly different:
Shown in Fig.~\ref{fig: Cubic Example 1}(b), the Poincaré map of
the flow shows that the origin is a saddle-type Lagrangian trajectory
with transversely intersecting stable (blue) and unstable (red) manifolds,
which lead to chaotic mixing near the origin. Therefore, the Navier--Stokes
solution (\ref{eq: Cubic example}) with $\alpha(t)\equiv0.005$ provides,
similarly to Example \ref{Example 2}, a false positive for coherent
material vortex detection based on streamlines and on the Okubo--Weiss
criterion. 
\end{example}

\begin{example}
We now consider another universal Navier--Stokes solution of the
form
\begin{widetext}
\begin{equation}
\mathbf{u}(\mathbf{x},t)=\left(\begin{array}{cc}
\sin4t & \cos4t+\frac{1}{2}\\
\cos4t-\frac{1}{2} & -\sin4t
\end{array}\right)\mathbf{x}+\alpha(t)\left(\begin{array}{c}
x\left(x^{2}-3y^{2}\right)\\
-y\left(3x^{2}-y^{2}\right)
\end{array}\right),\label{eq: Cubic example 2}
\end{equation}
\end{widetext}
where we have chosen $a_{2}(t)\equiv0$, $b_{2}(t)\equiv0$, $a_{3}(t)\equiv\alpha(t)$
and $b_{3}(t)\equiv0$ in the quadratic and cubic terms of (\ref{eq: Exact solution form}),
respectively, and selected $\mathbf{h}(t)$, $\omega$, $a_{k}(t)$
and $b_{k}(t)$, for $k=2$ and $k>3$ as in Example 1. As discussed
in Example 3, for the choice of a small parameter $\alpha(t)\equiv0.005$,
we expect, by the KAM theorem \cite{arnold1989mathematical}, that
most quasiperiodic motions linearized system survive around the origin.
Contrary to this, the instantaneous streamline picture, shown in Fig.~\ref{fig: Cubic example 2}(a)
for $t=0$, suggests a stagnation point at the origin. The Okubo--Weiss
criterion, visualized by the yellow domains ($OW<0$) in Fig.~\ref{fig: Cubic example 2}(a),
predicts two coherent vortices away from the origin. Fig.~\ref{fig: Cubic example 2}(b)
shows in blue KAM curves of the velocity field (\ref{eq: Cubic example 2}),
revealing a bounded material coherent vortex around the origin. The
rest of the particles, which are not captured by the material vortex,
escape to infinity. Hence the Okubo--Weiss criterion provides one
false negative and two false positives for vortex identification in
the velocity field (\ref{eq: Cubic example 2}).

\begin{figure*}
\includegraphics[viewport=100bp 90bp 1900bp 1000bp,scale=0.28]{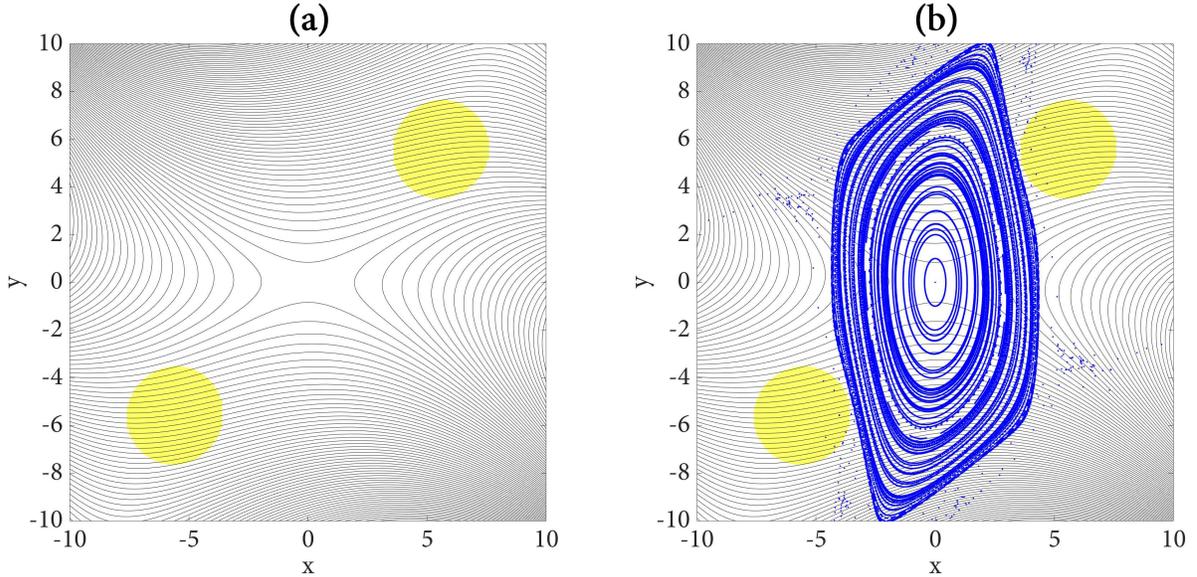}\caption{(a) Instantaneous streamlines and Okubo--Weiss elliptic region (yellow)
for the universal Navier--Stokes solution (\ref{eq: Cubic example 2})
with $\alpha(t)\equiv0.005$ at time $t=0$. Other time slices are
similar. (b) KAM curves (blue) of the same velocity field.\label{fig: Cubic example 2}}
\end{figure*}
\end{example}

\begin{example}
A further extension of the velocity field (\ref{eq: Cubic example 2})
is given by the universal Navier--Stokes solution 
\begin{widetext}
\begin{equation}
\mathbf{u}(\mathbf{x},t)=\left(\begin{array}{cc}
\sin10t & \cos10t+2\\
\cos10t-2 & -\mathrm{sin}10t
\end{array}\right)\mathbf{x}+\alpha(t)\left(\begin{array}{c}
x\left(x^{2}-3y^{2}\right)\\
-y\left(3x^{2}-y^{2}\right)
\end{array}\right),\label{eq: Cubic example 3}
\end{equation}
\end{widetext}
where we have chosen $a_{2}(t)\equiv0$, $b_{2}(t)\equiv0$, $a_{3}(t)\equiv\alpha(t)$
and $b_{3}(t)\equiv0$ in the quadratic and cubic terms of (\ref{eq: Exact solution form}),
respectively, and set $\mathbf{h}(t)$, $\alpha(t)$, $\omega$, $a_{k}(t)$
and $b_{k}(t)$ for $k>3$ as in Example 4. This solution is a nonlinear
extension of the general linear velocity field (\ref{eq:Generalized linear example}),
with $\omega=-2$ and $C=10$. Therefore, as for the solution (\ref{eq:Quadratic example 2}),
$|C-\omega|>2$. Hence the origin of (\ref{eq: Cubic example 3})
is a center-type fixed point of the linearized system. By the KAM
theorem \cite{arnold1989mathematical}, adding a nonlinear term multiplied
by the small parameter $\alpha(t)\equiv0.005$ in (\ref{eq: Cubic example 3}),
the Lagrangian particle motion in the cubic velocity field (\ref{eq: Cubic example 3})
is expected to remain elliptical (vortical) around the origin. The
instantaneous streamlines of (\ref{eq: Cubic example 3}), shown in
Fig.~\ref{fig: Cubic example 3}(a) for $t=0$, suggest a coherent
vortex around the origin. The Okubo--Weiss criterion, visualized
in Fig.~\ref{fig: Cubic example 3}(a) for the initial time $t=0$
by the yellow domain ($OW>0$), also suggests a coherent vortex near
the origin.
\begin{figure*}
\includegraphics[viewport=100bp 90bp 1900bp 1000bp,scale=0.28]{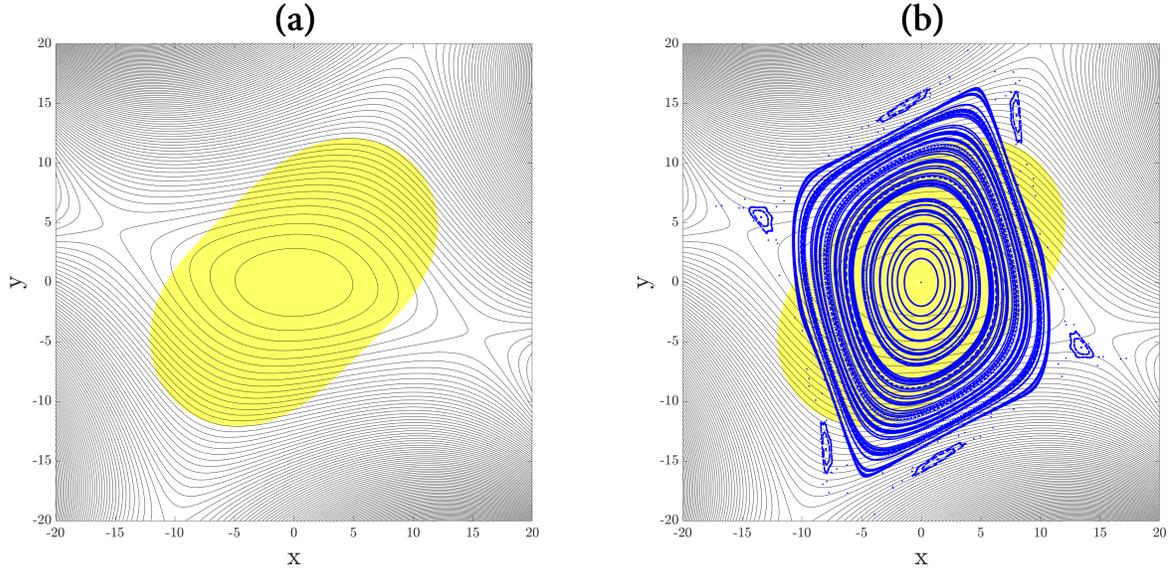}\caption{\label{fig:Cubic example 3}(a) Instantaneous streamlines and Okubo--Weiss
elliptic region (yellow) for the universal Navier--Stokes solution
(\ref{eq: Cubic example 3}) with $\alpha(t)\equiv0.005$ at time
$t=0$. Other time slices are similar. (b) KAM curves (blue) of the
same velocity field.\label{fig: Cubic example 3}}
\end{figure*}
Shown in blue in Fig.~\ref{fig: Cubic example 3}(b) are KAM curves
of the flow, revealing indeed a bounded coherent vortex around the
origin. The rest of the particles, which are not captured by the material
vortices around the origin, escape to infinity. Hence for the solution
(\ref{eq: Cubic example 3}), both the Okubo--Weiss criterion and
the instantaneous streamlines correctly predict the presence of a
coherent vortex around the origin. At the same time, they fail to
predict the correct shape of the vortex, and completely miss six smaller
vortices surrounding the large vortex in Fig.~\ref{fig: Cubic example 3}(b). 
\end{example}

\section{Conclusions}

We have derived an explicit form for all spatially polynomial, universal,
planar Navier--Stokes flows up to arbitrary order. We then used examples
of such solutions to test the ability of the instantaneous streamlines
and of the Okubo--Weiss criterion, the 2D version of the $Q$-criterion,
to detect coherent material vortices and stretching regions in unsteady
flows. 

Specifically, using the main result of this paper, we have derived
two chaotically mixing Navier--Stokes flows whose analysis via instantaneous
streamlines and by the Okubo--Weiss criterion suggests a lack of
stretching due to the presence of a coherent vortex. Likewise, we
have constructed two exact Navier--Stokes flows that have a bounded
coherent Lagrangian vortex around the origin despite the hyperbolic
flow structure suggested by instantaneous streamlines and the Okubo--Weiss
criterion. Finally, we have constructed a Navier--Stokes solution
whose trajectories form a coherent vortex near the origin. While the
Okubo--Weiss criterion and the instantaneous streamlines do signal
a nearby vortex in this example, they fail to render the correct shape
of the vortex and miss additional smaller vortices in its neighborhood. 

Using the 2D unsteady solutions, we have given an explicit family
of unsteady polynomial solutions to the 3D Navier--Stokes equation.
The first two coordinates of these 3D velocity fields agree with our
planar polynomial solutions, while their third coordinate is simply
a uniform, constant velocity component. When applied to these extended
unsteady solutions, the $Q$-, $\Delta$-, $\lambda_{2}$- and $\lambda_{ci}$-criteria
give the same incorrect flow classification results as the Okubo--Weiss
criterion does in our two-dimensional examples. 

The exact solutions derived in this paper can be used as basic unsteady
benchmarks for coherent structure detection criteria and numerical
schemes. They also provide a wealth of bounded, dynamically consistent
flow patterns away from boundaries. For instance, the specific two-dimensional
velocity field examples we have derived can be viewed as models of
coherent structures, such as eddies and fronts, in oceanic flows away
from the coastlines.

\begin{acknowledgments}
We would like to acknowledge useful conversations with Mattia Serra
on the subject of this paper. This work was partially supported by
the Turbulent Superstructures Program of the German National Science
Foundation (DFG). 
\end{acknowledgments}

\nocite{*}
\bibliography{aipsamp2}

\end{document}